# Perishable Goods versus Re-tradable Assets:

## A Theoretical Reappraisal of a Fundamental Dichotomy[1]

Sabiou M. Inoua and Vernon L. Smith[2]

Chapman University

**Abstract**. Experimental results on market behavior establish a lower stability and efficiency of markets for durable re-tradable assets compared to markets for non-durable, or perishable, goods. In this chapter, we revisit this known but underappreciated dichotomy of goods in the light of our theory of competitive market price formation, and we emphasize the fundamental nature of the concept of asset re-tradability in neoclassical finance through a simple reformulation of the famous no-trade and no-arbitrage theorems.

## 1 Stability of perishable final goods: experiments and a classical model

Market experiments conducted mid twentieth century (Chamberlin, 1948; Smith, 1962) established the stability, efficiency, and robustness of markets for perishable goods under notably the double-auction market institution (Smith, 1962). These laboratory results are now well-known (reviewed, e.g., in Plott, 1982; Smith, 1982; Smith & Williams, 1990), have been replicated many times around the world (Lin et al., 2020), and motivated a few neoclassical models of price equilibration in double-auction markets (Wilson, 1987; Friedman, 1991; Cason & Friedman, 1996; Gjerstad & Dickhaut, 1998; Anufriev et al., 2013; Asparouhova et al., 2020).[3] Alternatively, if one sacrifices some institutional details for a general principle, then one can characterize a single-market price mechanism as convergence to competitive equilibrium defined (more generally than to a market-clearing price) as a generalized median of traders' values and costs, treated as primitive concepts, and as an operational substitute for the utility function, following an old, forgotten, more or less explicit, deep classical methodological tradition of supply and demand (Inoua & Smith, 2021a, 2022).[4] That is, competitive price dynamics is rooted in the minimization of the following distance function, which measures the total potential surplus available to all buyers and sellers in a market at any standing transaction price (or equivalently the function measures the sum of mutually beneficial potential trades available in a market, that are not yet actualized, at any arbitrary standing price):

$$V(p) = \sum_{v \geq p} |v - p| + \sum_{c \leq p} |c - p|, \qquad (0.1)$$

where the notation means summation of all values $v \geq p$ and all costs $c \leq p$ (because only profitable units will be traded) and $p$ is a standing transaction price. That is, if a transaction price sequence $\{p_t : t = 1,...,T\}$ emerges from the competition of traders (in the sense of buyer-buyer outbidding, seller-seller underselling, and buyer-seller haggling) then it corresponds to a non-increasing sequence $\{V(p_t) : t = 1,...,T\}$ of the potential surplus function:

$$V(p_{t+1}) \leq V(p_t), \; t = 1, 2, ..., T. \qquad (0.2)$$

---

[1] An edited version to appear in the *Handbook of Experimental Finance*, Sascha Füllbrunn and Ernan Haruvy (eds), Edward Elgar Publishing. We thank the Editors for the invitation to contribute to the handbook. We also thank J. Huber, N.H. Nax, S. Lin, and D. Porter for their help in accessing experimental data.

[2] Economic Science Institute, Chapman University, 1 University Drive, Orange, CA 92866, USA; vsmith@chapman.edu; inoua@chapman.edu.

[3] There is also the "zero-intelligence" trader model (Gode & Sunder, 1993).

---

[4] This approach is based on the old classical view of competition as a collective haggling and bargaining process, which more or less explicitly is based on reservation prices as a primitive concept in a partial-equilibrium context (Inoua & Smith, 2022). (In the general equilibrium context, one replaces the concept of consumer's reservation price for a good with consumer wealth, viewed as the maximum the consumer would be willing to pay for the maximum number of units of all goods needed. We will not here elaborate on this point of the theory of classical price formation, the subject of ongoing research by the authors.) The minimum principle was already hinted in the seminal experimental paper in a different form and under a different name, "the excess-rent hypothesis" (Smith, 1962, Section V).

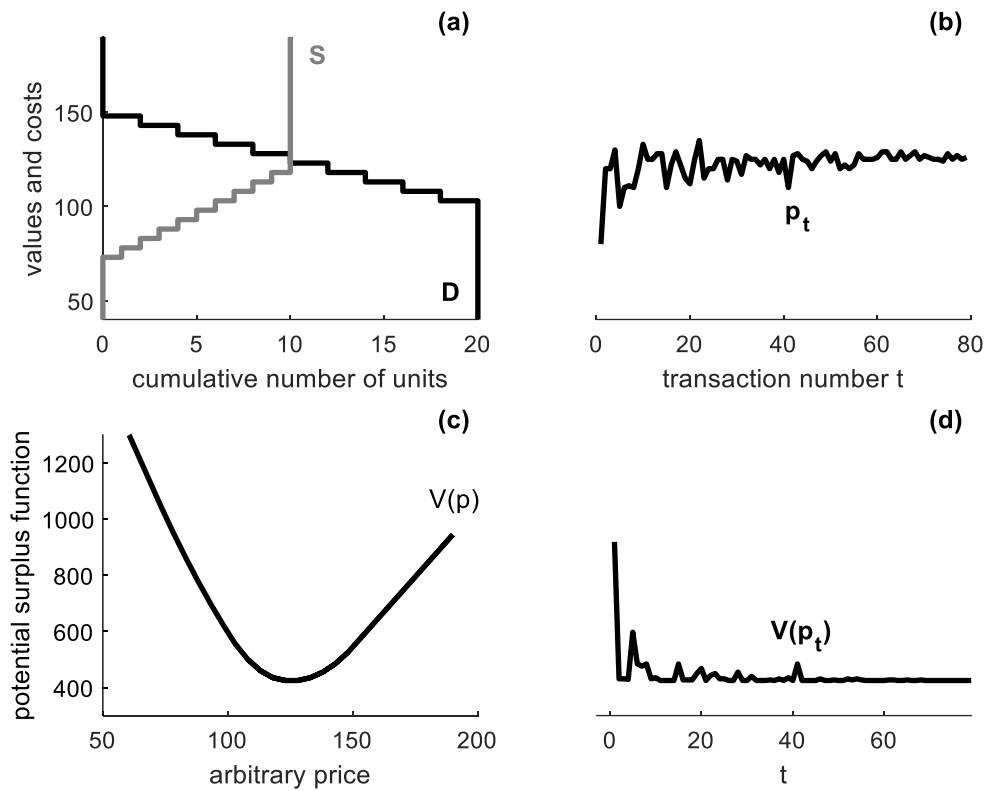

Figure 1. The Maximum Information Principle (PMI) Illustrated using Lab Data: (a) supply and demand (value and cost distributions), (b) transaction price dynamics; (c) potential surplus function V; dynamics of potential surplus function V.[5]

This minimum principle is an informational characterization of a competitive market because it says that a competitive price, which minimizes the potential surplus function, is a robust optimal summary of the traders' valuations. It is informational in the sense that contract prices publicize value-cost information that, pre-market, is inherently private and decentralized. Thus, we refer to it as the principle of maximum information (PMI).[6]

The price stability of a good traded for the satisfaction of its final consumer use-value, as in Figure 1, contrasts starkly with the prices observed in asset market experiments demonstrating the occurrence of asset price bubbles (Smith et al., 1988).[7] This contrast in stability arises because consumer goods have value only in use, which governs their market price. Any good durable enough to re-trade exhibits both a use value and a resale value; if such a good's resale price disconnects from its use-value price, the good may trade in a price bubble that deviates substantially, if unsustainably, from its fundamental use value. In the next section we articulate a theory of the "excess volatility" of re-tradable goods in terms of their being bought for the prospect of being resold for capital gains. For example, Figure 2 contrasts No Re-trade with Re-trade, as experimental treatment conditions under the same supply and demand configuration, demonstrating the effect of Re-trade on the price stability observed in No-Re-trade.

---

[5] Data source: Ikica et al. (2021, "FullMoreB" treatment).

[6] The principle is in fact more transparently stated, in an equivalent manner, in terms of an informational function a la Shannon (Inoua & Smith, 2021a, 2022).

Figure 2. The maximum information principle and its breakdown in the Presence of Re-trade and Excess Liquidity. Standard convergence to competitive equilibrium holds when No-re-trade is allowed (Treatment "P2-SP1", Column 2); but stability is lost (counteracted

---

[7] For a review of bubble experiments, see Palan (2013).

by speculation) when units can be re-traded (Treatment "P2-RT1), and the instability is greater when subject had greater cash endowment for Re-trade (Treatment "P1-RT1").[8]

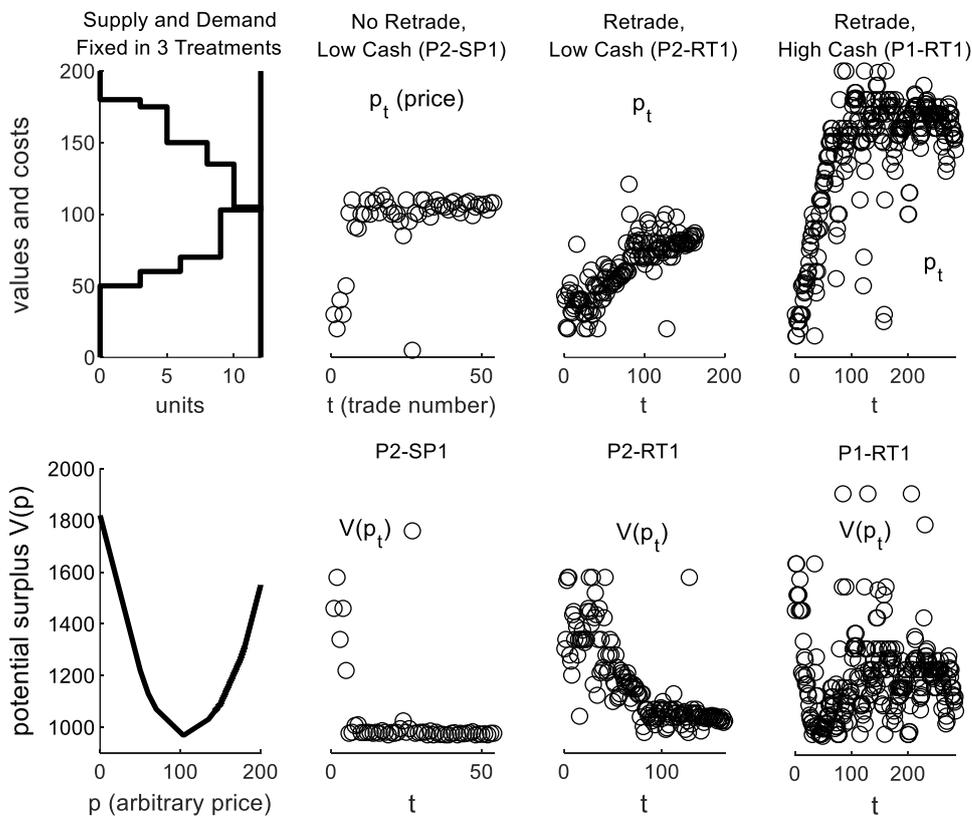

## 2 Speculation and excess volatility of re-tradable assets[9]

Consider a re-tradable (durable) asset exchanged in a market in which all the participants are (long-run) investors, i.e. they trade the asset based on their subjective evaluations $\{v\}$ of information on the asset's fundamental value as reflecting the real economic prospect of the asset's issuer, and which is therefore *exogenous to the market*, namely to the price dynamics; then the minimum principle says that the competitive asset price is a median of the traders' valuations of the asset. Further, it is then reasonable to expect that a core tenet of the efficient market hypothesis (Fama, 1970) will hold in the market thus modeled, because the competitive price will approximate the asset's real value (as long as the subjective traders' valuations are not biased *in the aggregate*). It is otherwise, however, when trend-following speculation enters the picture.

---

[8] Data source: Dickhaut et al. (2012).

[9] For a more detailed exposition of some of the ideas presented briefly in Sections 2 and 3, see Inoua and Smith (2021b), to appear in the *Journal of Behavioral and Experimental Finance*.

Convergence to a fixed equilibrium point is of course at odds with the dynamics of speculative prices, which are prone to extreme (non-Gaussian) fluctuations. More precisely, speculative price changes are power-law distributed, as is known since Mandelbrot's seminal finding (Fama, 1963; Mandelbrot, 1963b, 1963a; Gopikrishnan et al., 1998; Plerou et al., 2006; Bouchaud, 2011):[10] Thus,

$$\text{prob}\{|r| \geq x\} \sim \frac{C}{x^\alpha}, \tag{0.3}$$

for large $x$, where $r = \Delta p/p$ is the asset's percent return (relative price change), the exponent $\alpha$ is typically close to 3, and $C$ is just a normalizing constant. A second universal regularity (which we do not discuss in greater detail here) is volatility clustering: large price changes tend to be clustered in time (i.e., small-magnitude price changes tend to be followed by small-magnitude price changes, and large-magnitude price changes by large-magnitude price changes): formally, while the return process is serially uncorrelated, its magnitude (or absolute value) is long-range correlated. These empirical regularities have also been observed in the lab (Plott & Sunder, 1982) and have been closely investigated experimentally (Kirchler & Huber, 2007, 2009).

Figure 3. Power-law and clustered volatility illustrated: General Electric stock. (a) Price. (b) Return (in percent). (c) Cumulative distribution of volatility in log-log scale, and a linear fit of the tail, with a slope close to 3; (d) Autocorrelation function of

---

[10] The notation $f(x) \sim g(x)$ means $f(x)/g(x) \to 1$ as $x \to \infty$.

return, which is almost zero at all lags, while that of volatility is nonzero over a long range of lags (a phenomenon known as volatility clustering).[11]

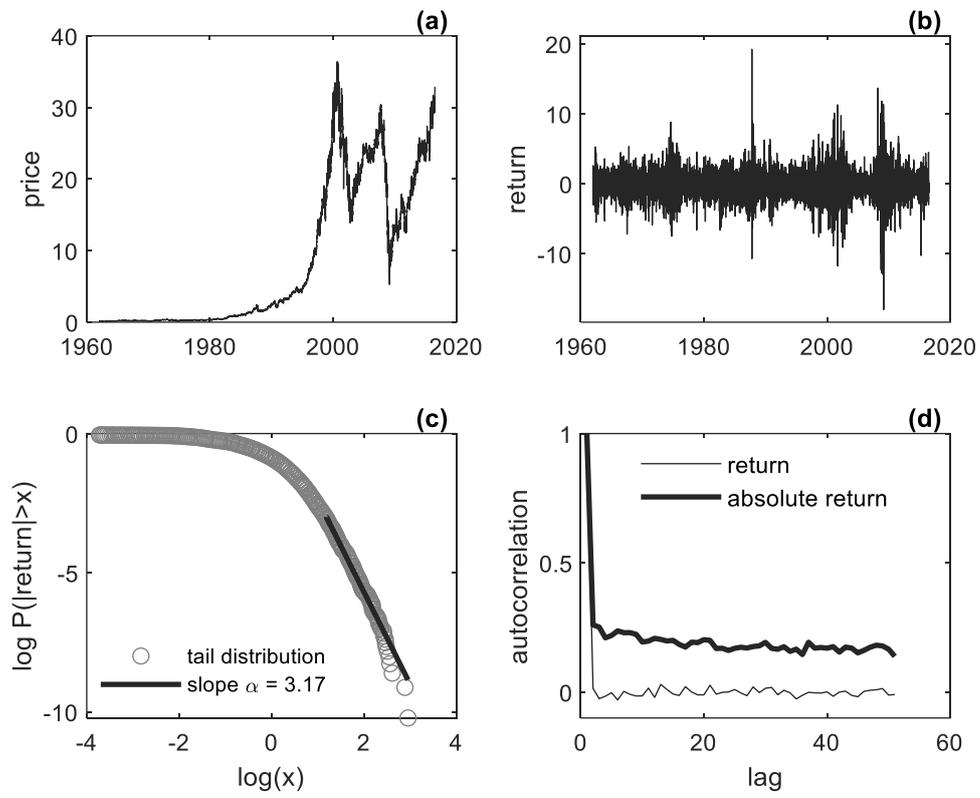

Figure 4. Price Volatility in the High-Cash-Re-trade Treatment in Figure 2 (third column, but here all periods combined for statistical significance of the estimation of the power law tail exponent).

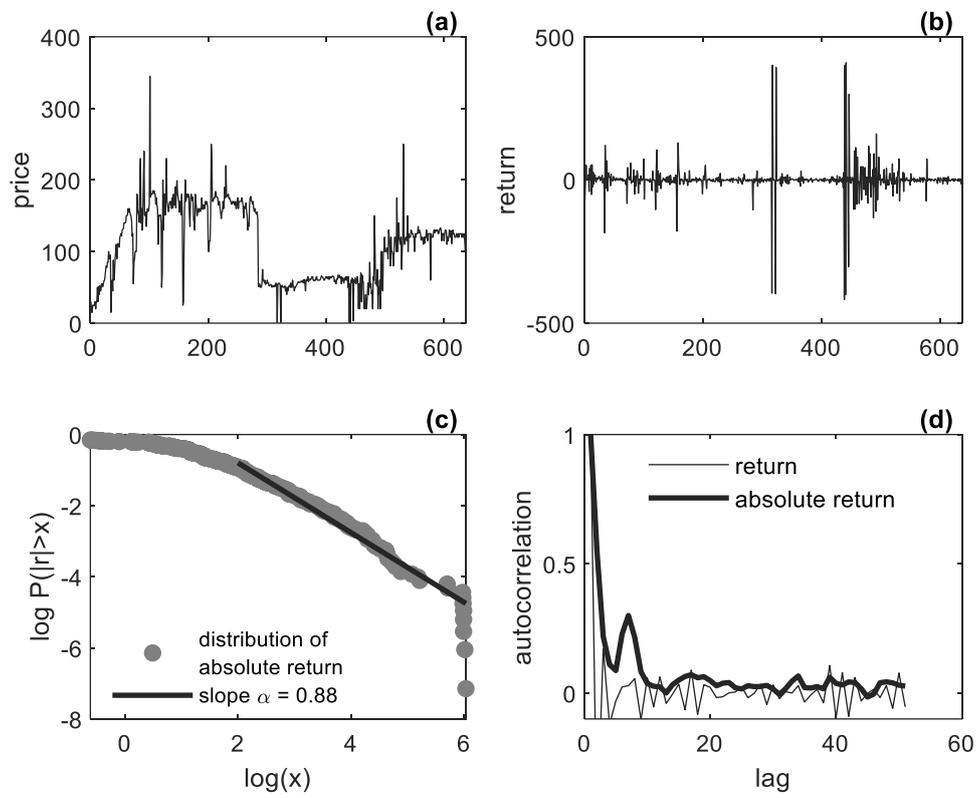

Figure 5. A Lab asset price volatility.[12]

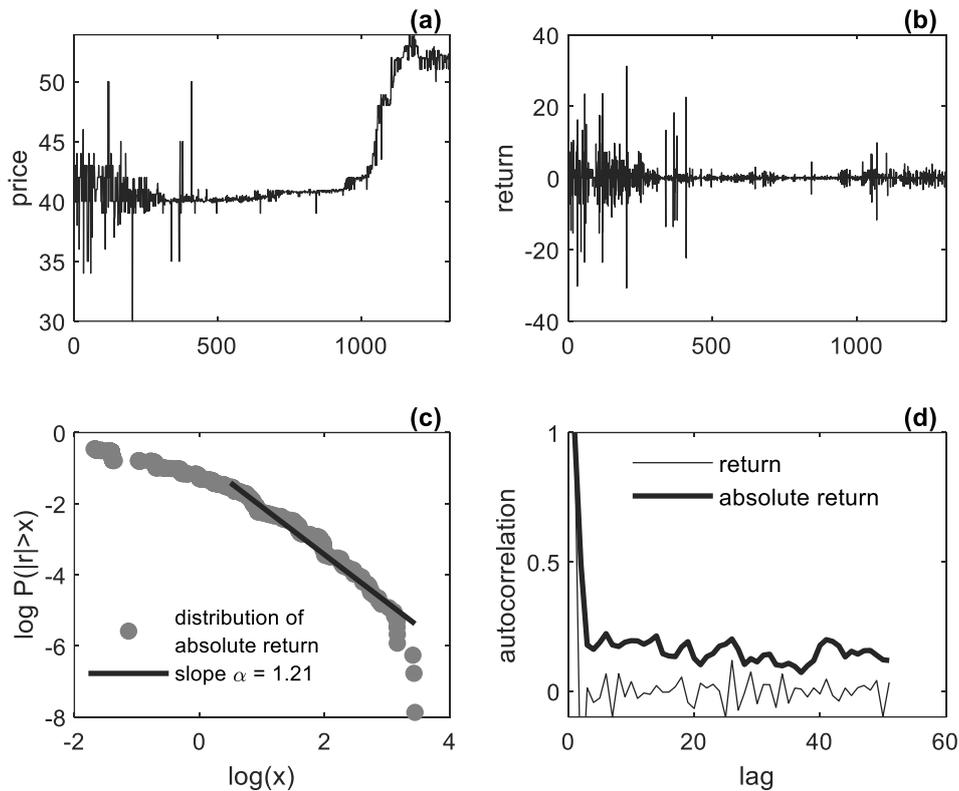

While the lab asset price volatility are clearly clustered and fat-tailed (non-Gaussian) upon graphical inspection, the price series (even when pooled across periods) are typically not long enough to allow for a fully comfortable statistical analysis of the tails of the price series (fat tails require a much longer data series for statistical significance than common distributions).[13] Notwithstanding this caveat, we observe that the lab asset price volatility is typically more extreme than its field counterpart (Figure 3 versus Figure 4 and Figure 5: the lower the tail exponent $\alpha$ of a power law, the higher the volatility).

---

[12] Data source: Kirchler and Huber (2009, Market 5).

[13] While it is not unreasonable to pool data across periods for a fixed treatment as the data-generating process can be reasonably assumed stationarity, it is otherwise, however, for pooling across different treatments when the parameters change greatly.) Finally, note that volatility clustering cannot be rigorously defined in terms of an autocorrelation function when the power law exponent is less than or equal to 2, for then the return process has an infinite-variance process.

There is as yet no consensus among experts as to a canonical explanation of the emergence of the power law of asset returns (and of volatility clustering), although theoretical models abound in this field, notably the interesting, if often mathematically intractable, agent-based models (reviewed, e.g., by Lux & Alfarano, 2016). Here we suggest a simple explanation found in the continuity of the previous characterization of the price mechanism.

The speculator who expects a price increase of the asset, buys; the one who expects a price drop sells: thus by construction, a speculator's effective reservation price for a good is his anticipated future resale price, say $p^e$ (or more precisely the minimum between the anticipated resale price and the maximum amount the speculator could pay for the unit given the speculator *liquidity* constraint, an important aspect not discussed here). Thus, the potential surplus function (0.1) becomes, for a purely speculative market:

$$V(p,\{p^e\}) = \sum_{\{p^e\}} |p^e - p|. \tag{0.4}$$

where we sum over the distribution of traders' anticipated resale prices. There is no reason why the speculative version (0.4) of the potential surplus function would be minimized (and that the market would converge to a fixed competitive equilibrium), for the median anticipated resale price need not be given if the traders' expectations are self-reinforcing, which is the case if speculators follow short-run price trends, as they do in practice. One can easily show that the price change of an asset traded in a competitive market populated by trend-following speculators, follows (at a first-order linear approximation) a random-coefficient autoregressive model (Inoua, 2020; Inoua & Smith, 2021b):

$$r_t = \sum_{h=1}^{H} \alpha_{ht} r_{t-h} + \varepsilon_t, \tag{0.5}$$

where $\{\alpha_{ht}\}$ and $\{\varepsilon_t\}$ are random variables. Random-coefficient autoregressive processes (also known as Kesten processes) are rigorously studied by mathematicians (Kesten, 1973; Klüppelberg & Pergamenchtchikov, 2004; Buraczewski et al., 2016) and are perhaps the most natural class of power-law generating processes, where the tail exponent depends on the distribution of the feedback coefficients $\{\alpha_h\}$. But they cannot generate clustered volatility, which can be explained simply in terms of traders' reaction to exogenous news about the economy (Inoua, 2020; Inoua & Smith, 2021b).

Figure 6. A purely speculative asset market model: (a) price; (b) asset return follows a first-order random-coefficient auto-regressive process.; (c) power-law distribution of asset return; (d) autocorrelation functions of return and absolute return.

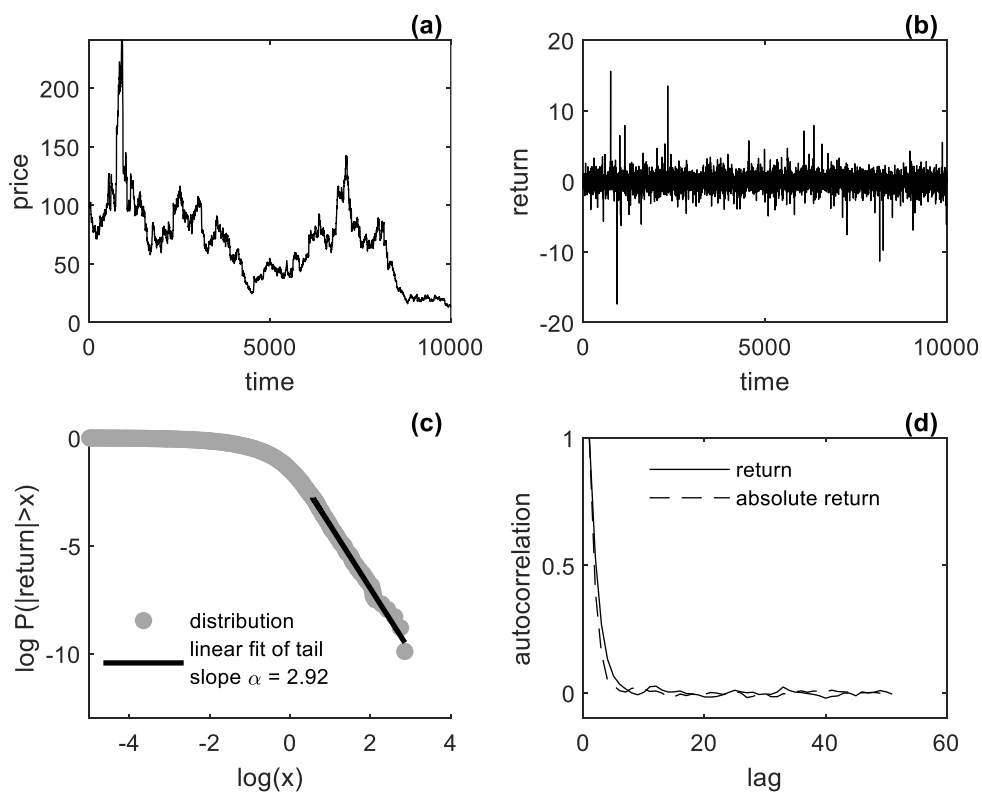

This model of financial volatility rests on assumptions rooted in common practices in finance (speculation, trend following, reaction to news), which are hard to articulate, however, in a neoclassical framework owing to the no-trade and no-speculation theorems (Rubinstein,

1975; Milgrom & Stokey, 1982; Tirole, 1982). We show that this inherent limitation of neoclassical finance can be formulated in terms of the concept of re-tradable asset.

**3 Re-tradability and the no-trade and no-arbitrage theorems**

An arbitrage-free market, recall, is one in which an asset's price is the discounted expected value of the asset's future payoff:

$$p_t = \mathbb{E}_t(p_{t+1} + d_{t+1}), \qquad (0.6)$$

where $p_t$ and $d_t$ are respectively the asset's price and dividend payoff per unit asset holding at the closing of period $t$, and $\mathbb{E}_t$ is the discounted expectation operator (where the discount factor is strictly positive) conditional on available information about price history and dividend announcement up to the end of period $t$. The standard no-arbitrage asset price formula (0.6) is equivalent to a no-re-trade theorem. The argument is simple:

Consider a financial market in which multiple assets can be traded at exogenous, given, prices $\mathbf{p}$. (From now on, boldface denotes a vector.) For each period $t$, a trader's financial wealth $W_t$ (evaluated at the end of each period $t$) is the market value of his asset holdings $\mathbf{H}_t$, minus his asset purchase cost, plus his resale revenue, plus his dividends received, during that period: $W_t = \mathbf{p}_t \cdot \mathbf{H}_t - \Delta \mathbf{H}_t \cdot \mathbf{p}_{t-1} + \mathbf{d}_t \cdot \mathbf{H}_t$, $t \geq 1$, starting from an initial wealth $W_0 \equiv \mathbf{p}_0 \cdot \mathbf{H}_0$ (the value of the trader's initial asset holdings), and $\Delta \mathbf{H}_t$ being the traders' transaction during period $t$, decided based on available information (up to the end of period $t-1$ (just for notational simplicity, we assume by convention that all trades during period $t$ are to be executed at $\mathbf{p}_{t-1}$, the prices announced at the closing of period $t-1$). Had the trader maintained his asset position throughout period $t$, refraining from re-trading any unit

($\Delta \mathbf{H}_t = 0$), he would enjoy a wealth of $W_t^* = \mathbf{p}_t \cdot \mathbf{H}_{t-1} + \mathbf{d}_t \cdot \mathbf{H}_{t-1}$. The relative advantage of re-trading assets over holding one's position is measured by $R_t = W_t - W_t^*$, namely:

$$R_t = (\mathbf{p}_t - \mathbf{p}_{t-1} + \mathbf{d}_t) \cdot \Delta \mathbf{H}_t. \tag{0.7}$$

In an arbitrage-free market, $\mathbb{E}_t(R_t) = 0$.[14] Hence no risk-averse expected-utility maximizer would re-trade any asset holdings in an arbitrage-free market (by Jensen's inequality). That the no-arbitrage condition (0.6) has a straightforward formulation in terms of the concept of re-trade advantage (0.7) offers in fact a natural way of formulating the mathematics of arbitrage-free markets, whose standard formulation (Ross, 1976; Rubinstein, 1976; Ross, 1978; Harrison & Kreps, 1979; Harrison & Pliska, 1981; Dalang et al., 1990) appeals instead to the concept of nominal capital gains from a self-financed portfolio.

---

[14] This is true by definition of arbitrage-free market [equation (0.6)] and by the fact that a trader's decision to trade during period *t* is based on information available up to the end of period $t-1$, so that $\Delta \mathbf{H}_t$ is known (determinate, constant) given information available at the end of period *t*. (In technical terms, the re-trade decisions $\{\Delta \mathbf{H}_t\}$ form a "predictable process" with respect to information available up to the end of period *t*.)